\documentclass[aps,twocolumn,floatfix, groupedaddress, superscriptaddress]{revtex4-1}
\usepackage{bm}
\usepackage{dcolumn}
\usepackage{graphicx}
\usepackage{amsmath}
\usepackage{amssymb}
\usepackage{amsfonts}
\usepackage{float}
\usepackage{multirow}
\usepackage{xcolor}
\usepackage[capitalise]{cleveref}

\begin{document}
\title{Electronic structure of biased alternating-twist multilayer graphene} 
\author{Kyungjin Shin}
\thanks{These authors contributed equally to this work.}
\affiliation{Department of Physics and Astronomy, Seoul National University, Seoul 08826, Korea}
\author{Yunsu Jang}
\thanks{These authors contributed equally to this work.}
\affiliation{Department of Physics and Astronomy, Seoul National University, Seoul 08826, Korea}
\author{Jiseon Shin}
\affiliation{Department of Physics, University of Seoul, Seoul 02504, Korea}
\affiliation{LG Electronics, CTO Division, Seocho R\&D Campus, Seoul 06772, Korea}
\author{Jeil Jung}
\email{jeiljung@uos.ac.kr}
\affiliation{Department of Physics, University of Seoul, Seoul 02504, Korea}
\affiliation{Department of Smart Cities, University of Seoul, Seoul 02504, Korea}
\author{Hongki Min}
\email{hmin@snu.ac.kr}
\affiliation{Department of Physics and Astronomy, Seoul National University, Seoul 08826, Korea}

\date{\today}

\begin{abstract}
We theoretically study the energy and optical absorption spectra of alternating twist multilayer graphene (ATMG) under a perpendicular electric field.
We obtain analytically the low-energy effective Hamiltonian of ATMG up to pentalayer in the presence of the interlayer bias by means of first-order degenerate-state perturbation theory, and present general rules for constructing the effective Hamiltonian for an arbitrary number of layers.
Our analytical results agree to an excellent degree of accuracy with the numerical calculations for twist angles $\theta \gtrsim 2.2^{\circ}$ that are larger than the typical range of magic angles.
We also calculate the optical conductivity of ATMG and determine its characteristic optical spectrum, which is tunable by the interlayer bias.
When the interlayer potential difference is applied between consecutive layers of ATMG, the Dirac cones at the two moir\'{e} Brillouin zone corners $\bar{K}$ and $\bar{K}'$ acquire different Fermi velocities, generally smaller than that of monolayer graphene, and the cones split proportionally in energy resulting in a step-like feature in the optical conductivity.
\end{abstract}

\maketitle

\section{Introduction}
Twisted graphene systems have attracted widespread attention after the discovery of superconductivity and correlated insulating states~\cite{Cao2018a,Cao2018b, Yankowitz2019,Lu2019} in magic-angle twisted bilayer graphene (TBG).
By twisting two graphene layers, a new long-period structure, called a moir\'{e} superlattice, emerges due to spatially varying interlayer coupling, generating a unique band structure and associated electronic properties which strongly depend on the twist angle. Especially at the so-called magic angles, the Fermi velocity vanishes and nearly flat bands are formed~\cite{Laissardiere2010, Morell2010,Bistritzer2011,Tarnopolsky2019}, providing an ideal platform to study correlated electron phenomena where electron-electron interactions are dominant over the kinetic energy.

Studies beyond TBG have been extended to systems like twisted double-bilayer graphene~\cite{Koshino2019,Chebrolu2019,Lee2019,Shen2020,Liu2020,He2021} and twisted triple-bilayer graphene~\cite{Shin2022}, and even to other two-dimensional moir\'{e} material systems~\cite{Naik2018,Kariyado2019,Xian2019,Chen2019,Arora2020,Wang2020,An2020}, that revealed interesting interaction-driven phenomena such as correlated insulating~\cite{Lee2019,Shen2020,Liu2020,He2021,Xian2019} and topological~\cite{Koshino2019} phases that are {\it in situ} tunable.

Among them, the alternating twist multilayer graphene (ATMG) has been studied intensively both theoretically~\cite{Khalaf2019,Carr2020,Bezanilla2020,Lei2021,Phong2021,Lake2021,Choi2021,Nguyen2022,Leconte2022} and experimentally~\cite{Park2021,Hao2021,Cao2021,Kim2022,Park2022,Zhang2022,Shen2023} 
whose larger magic angle gives them an advantage over TBG. In particular, ATMG has attracted much attention due to its robust superconductivity observed from bilayer to pentalayer samples~\cite{Cao2018b,Yankowitz2019, Lu2019,Park2021,Hao2021,Cao2021,Kim2022,Park2022,Zhang2022}, while reports of superconductivity in other two-dimensional moir\'{e} systems are regarded as ambiguous~\cite{Liu2020,He2021,Chen2019,Arora2020,Wang2020,An2020,Balents2020}.

In this paper, we theoretically analyze the effect in the electronic structure of a perpendicular electric field on ATMG. Using first-order degenerate-state perturbation theory treating the interlayer potential as a perturbation, we analytically investigate the low-energy effective Hamiltonian and its energy spectrum that becomes more accurate as the twist angle is increased.
Then, we calculate the optical conductivity of biased ATMG that reveals a step-like feature arising from the splitting of Dirac nodes and their Fermi velocity renormalization introduced by the applied electric field.

The paper is organized as follows. In Sec.~\ref{Sec:Sec2}, we introduce a model of ATMG in the presence of the interlayer potential difference between consecutive layers and derive the corresponding low-energy effective Hamiltonian analytically up to pentalayer. We also present general rules for constructing the effective Hamiltonian of biased ATMG with an arbitrary number of layers.
In Sec.~\ref{Sec:Sec3}, we calculate the longitudinal optical conductivity of biased ATMG, and explain their characteristic optical absorption spectrum.
Finally, in Sec.~\ref{Sec:Sec4}, 
we discuss the interlayer coupling strength range for which our model is valid
and summarize our main results.

\section{Electronic structure}
\label{Sec:Sec2}
\subsection{Model}

\begin{figure}[htb]
\includegraphics[width=1.0\linewidth]{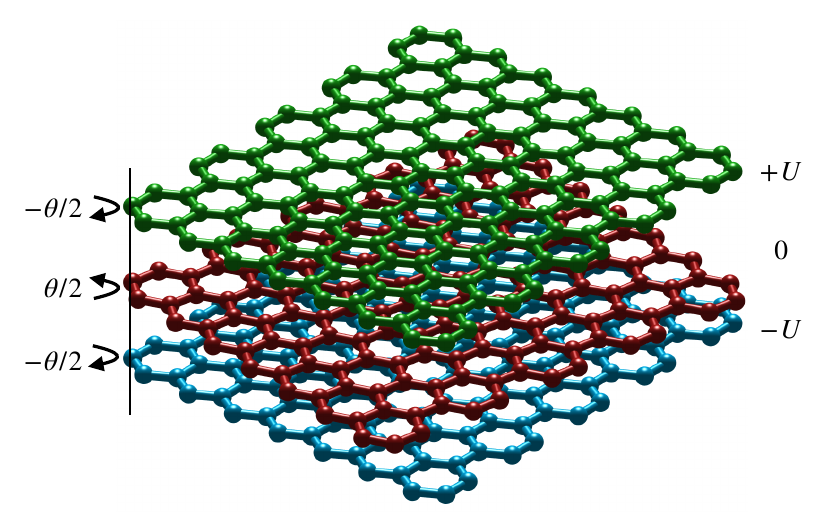}
\vspace{-5mm}
\caption{
Schematic illustration of the biased alternating twist multilayer graphene with $N=3$ layers.
}
\label{fig:fig1}
\end{figure}

We consider a model of vertically stacked $N$ graphene layers with the $\ell$th layer alternatingly twisted by an angle $\theta_{\ell}=(-1)^{\ell}\theta/2$, as shown in Fig.~\ref{fig:fig1}. For a perpendicular electric field, we assume that the interlayer potential difference $U$ is the same between the two adjacent layers. Following Leconte {\it et al.}~\cite{Leconte2022}, the Hamiltonian of ATMG in the presence of an interlayer potential difference can be expressed as
\begin{equation}
\label{eq:modelH}
H = 
\begin{pmatrix}
H_{\boldsymbol{k}}^{(1)} & T(\boldsymbol{r}) & 0 & \cdots\\
T^{\dagger}(\boldsymbol{r}) & H_{\boldsymbol{k}}^{(2)} & T^{\dagger}(\boldsymbol{r}) & \cdots\\
0 & T(\boldsymbol{r}) & H_{\boldsymbol{k}}^{(3)} & \cdots\\
\vdots & \vdots & \vdots & \ddots
\end{pmatrix} + V
\end{equation}
where the diagonal blocks $H_{\boldsymbol{k}}^{(\ell)}=\hbar v_{0} (\boldsymbol{k}\cdot\boldsymbol{\sigma}_{\theta_{\ell}})$ with $\boldsymbol{\sigma}_{\theta_{\ell}}=e^{\frac{i}{2}\theta_{\ell}\sigma_{z}}\boldsymbol{\sigma}e^{-\frac{i}{2}\theta_{\ell}\sigma_{z}}$ contain the Dirac cones of twisted graphene layers, $T(\boldsymbol{r})$ is the interlayer tunneling matrix, and $V={\rm {diag}}(V^{(1)}\mathbb{I}_{2}, V^{(2)}\mathbb{I}_{2}, ..., V^{(N)}\mathbb{I}_{2})$ is a diagonal matrix that captures the interlayer potential differences.
The Fermi velocity of the monolayer graphene is set as 
$v_{0}=\sqrt{3}a\lvert t \rvert/2\hbar\simeq10^{6}$~m/s with the lattice constant $a=2.46$~{$\rm \AA$} and the nearest-neighbor intralayer hopping parameter $t=-3.1$~eV.

In our model, the electric potential sequence $V^{(\ell)}$ is defined to satisfy both $V^{(\ell+1)}-V^{(\ell)}=U$ and $V^{(1)}=-V^{(N)}$, and the interlayer tunneling at the twisted interface takes the form
\begin{equation}
T(\boldsymbol{r}) = \sum_{j=0,\pm} e^{i\boldsymbol{q}_{j}\cdot\boldsymbol{r}}T^{j},
\end{equation}
where $\boldsymbol{q}_{0}$, $\boldsymbol{q}_{\pm}$ are given by $\boldsymbol{q}_{0}=2k_{\rm D}\sin(\theta/2)(0, -1)$, $\boldsymbol{q}_{\pm}=2k_{\rm D}\sin(\theta/2)(\pm\sqrt{3}/2, 1/2)$ with the Dirac momentum for monolayer graphene $k_{\rm D}=4\pi/3a$.
Our model also considers the corrugated lattice structure due to the effect of out-of-plane relaxation, leading to the larger interlayer spacing in AA-stacking region than the AB/BA-stacking region~\cite{Uchida2014,Wijk2015}, thus resulting in unequal intra/inter sublattice hopping terms $w$ and $w'$, respectively. Following the convention of an initial AA stacking configuration \cite{Jung2014}, the interlayer tunneling matrices are given by
\begin{equation}
T^{0} = 
\begin{pmatrix} 
w' & w \\
w & w' 
\end{pmatrix},\;T^{\pm} =
\begin{pmatrix} 
w' & w e^{\mp i2\pi/3} \\
w e^{\pm i2\pi/3} & w' 
\end{pmatrix},
\end{equation}
where $w'=0.0939$~eV and $w=0.12$~eV~\cite{Chebrolu2019}. For the full numerical calculations, we include the lattice corrugation $(w\neq w')$, whereas we assume the rigid model of equal hopping terms ($w=w'=0.12$~eV) for simplicity when we study the Hamiltonian analytically.
We choose representative twisted angles $3^{\circ}$$-$$5^{\circ}$ for the numerical calculations above the typical magic angle values that lie between $1^{\circ}$$-$$2^{\circ}$.
For these large angles the interlayer coupling substantially reduces the Fermi velocity of the dispersive Dirac cones near the two moir\'{e} Brillouin zone (mBZ) corners $\bar{K}$ and $\bar{K}'$ but have not completely flattened them.

In the absence of the interlayer potential difference, the effective Hamiltonian of the ATMG at $\bar{K}$ and $\bar{K}'$ can be described as a set of TBG models at different angles with an additional decoupled monolayer graphene model at $\bar{K}$ (or at $\bar{K}'$ depending on the continuum model we start with) for an odd number of layers~\cite{Khalaf2019}. The electronic structure of ATMG has a close analogy with Bernal-stacked multilayer graphene where the effective Hamiltonian is described by a set of bilayer graphene models with different effective masses with an additional decoupled monolayer graphene model for an odd number of layers~\cite{min2008a,min2008b}. This analogy between Bernal stacked multilayer graphene and ATMG can be expanded to their wave functions.

We now construct the wave function of ATMG at $\bar{K}$ or $\bar{K}'$ using the first shell model, in which the momentum-space lattice is truncated at the nearest-neighbor shell of the moir\'{e} reciprocal lattice $\bm{G}$ vectors, assuming the rigid model ($w = w'$). For the bilayer case ($N=2$), the zero-energy eigenstates near $\bar{K}$ and $\bar{K}'$ consist of four two-component spinors as follows:
\begin{equation}
\psi^{\rm{TBG}}_{\lambda, \bar{K} (\rm{or}\;\it{\bar{K}}')} = \frac{1}{\sqrt{1+6\alpha^{2}}}
\begin{pmatrix} 
a_{\lambda}\\
b_{\lambda}
\end{pmatrix}
\;\rm{or}\;
\begin{pmatrix} 
b_{\lambda}\\
a_{\lambda}
\end{pmatrix},
\label{eq:Psi_r}
\end{equation}
where $\alpha=w/\left[2v_{0}k_{\rm D}\sin(\theta/2)\right]$ is a dimensionless parameter. Following Bistritzer and MacDonald~\cite{Bistritzer2011}, we define $a_{\lambda}$ as a normalized eigenstate of $\hat{\boldsymbol{k}} \cdot \boldsymbol{\sigma}_{\theta_{\ell}}$ corresponding to the eigenvalue $\lambda=\pm 1$, and $b_{\lambda}=(b_{\boldsymbol{q}_{0}, \lambda}, b_{\boldsymbol{q}_{+}, \lambda}, b_{\boldsymbol{q}_{-}, \lambda})^{\rm T}$ determined by the equation $b_{\boldsymbol{q}_{j}, \lambda}=-h_{j}^{-1}T_{j}^{\dagger}a_{\lambda}$ with $h_{j}=\hbar v_{0}(\boldsymbol{k}+\boldsymbol{q}_{j}) \cdot \boldsymbol{\sigma}_{\theta_{\ell}}$. In a similar way to Bernal stacked multilayer graphene, the eigenfunctions of ATMG have the form of the solution of a one-dimensional chain problem, thus we can construct the eigenfunctions of ATMG in the following manner~\cite{Khalaf2019}:
\begin{equation}
\Psi_{r, \lambda}^{(\ell)} = \sqrt{\frac{2\tau}{N+1}}\;\sin(\ell\theta_{r})\;\psi^{\rm{TBG}}_{r, \lambda},
\label{eq:waveftn}
\end{equation}
where $\tau=2-\delta_{r,n+1}$, $\theta_{r} = r\pi/(N+1)$ with $r=1,2,\ldots,n$ for even $N=2n$ or for odd $N=2n+1$ with additional $r=(n+1)$th mode near $\bar{K}$, and $\psi^{\rm{TBG}}_{r, \lambda}$ can be obtained from $\psi^{\rm{TBG}}_{\lambda}$ in Eq.~(\ref{eq:Psi_r}) by letting $\alpha\rightarrow t_{r}\alpha$ and $b_{\lambda}\rightarrow t_{r}b_{\lambda}$ with $t_{r} = 2\,\cos{\theta_{r}}$.
Here, $\Psi_{r, \lambda}=(\Psi_{r, \lambda}^{(1)}, \Psi_{r, \lambda}^{(2)},\ldots,\Psi_{r, \lambda}^{(N)})^{\rm T}$ is a normalized eigenstate of the effective Hamiltonian $H_{\rm eff}=\hbar v_{r} (\boldsymbol{k}\cdot\boldsymbol{\sigma})$ of ATMG with $\lvert \Psi_{r, \lambda} \rvert^2=1$, where $v_{r}$ is a Fermi velocity of the Dirac cone $\Psi_{r, \lambda}$ given by
\begin{equation}
\frac{v_{r}}{v_{0}} = \frac{1-3t_{r}^{2}\alpha^{2}}{1+6t_{r}^{2}\alpha^{2}}.
\label{eq:velocity}
\end{equation}
Inserting $r=n+1$ in Eq.~(\ref{eq:velocity}), one finds that the $(n+1)$th mode for the odd number of layers corresponds to an eigenstate of the decoupled monolayer Hamiltonian.

In the presence of the interlayer potential difference $U$, we obtain analytically the low-energy effective Hamiltonian using first-order degenerate-state perturbation theory based on the minimal size Hamiltonian including only the first shell of the moir\'{e} $\bm{G}$ vectors. Due to the electric field, the Dirac cones near $\bar{K}$ or $\bar{K}'$ are hybridized and split from one another, so the effective Hamiltonian of each Dirac node would be altered in the following form:
\begin{equation}
H_{\rm eff} = \Delta(\alpha, U) + \hbar v^{*}(\boldsymbol{k}\cdot\boldsymbol{\sigma}),
\label{eq:H_eff}
\end{equation}
where $\Delta(\alpha, U)$ and $v^{*}$ are the energy shift and modified Fermi velocity of the effective Hamiltonian, respectively.
The energy shift due to the interlayer potential $U$ can be expressed as $\Delta(\alpha, U) = C(\alpha)U$, and the Fermi velocity $v^{*}$ can be expressed as a linear combination of $v_{r}$.
In the following Secs.~\ref{Sec:Sec2}.B and~\ref{Sec:Sec2}.C, we illustrate the effective Hamiltonian of $N=3$ and $4$ ATMG as examples. We leave the discussions of the analytical results for the $N=5$ case to Appendix~\ref{Sec:N=5}.

\subsection{$N=3$}
\label{subSec:N=3}

Here we derive the effective Hamiltonian of alternating twist trilayer graphene (AT3G) at the $\bar{K}$ and $\bar{K}'$ points of the mBZ. There are two Dirac cones centered at $\bar{K}$ with $v_{0}$ and $v_{1}$ Fermi velocities, as shown in Fig.~\ref{fig:fig2}(a), thus the size of the perturbation matrix $V_{\bar{K}}$ would be $2\times2$. Note that $v_{0}$ represents the Fermi velocity of monolayer graphene. Using Eq.~(\ref{eq:Psi_r}), we obtain the following normalized wave functions $\Psi_{r,\lambda}$ near $\bar{K}$ in our first shell model:

\begin{figure}[t]
\includegraphics[width=1.0\linewidth]{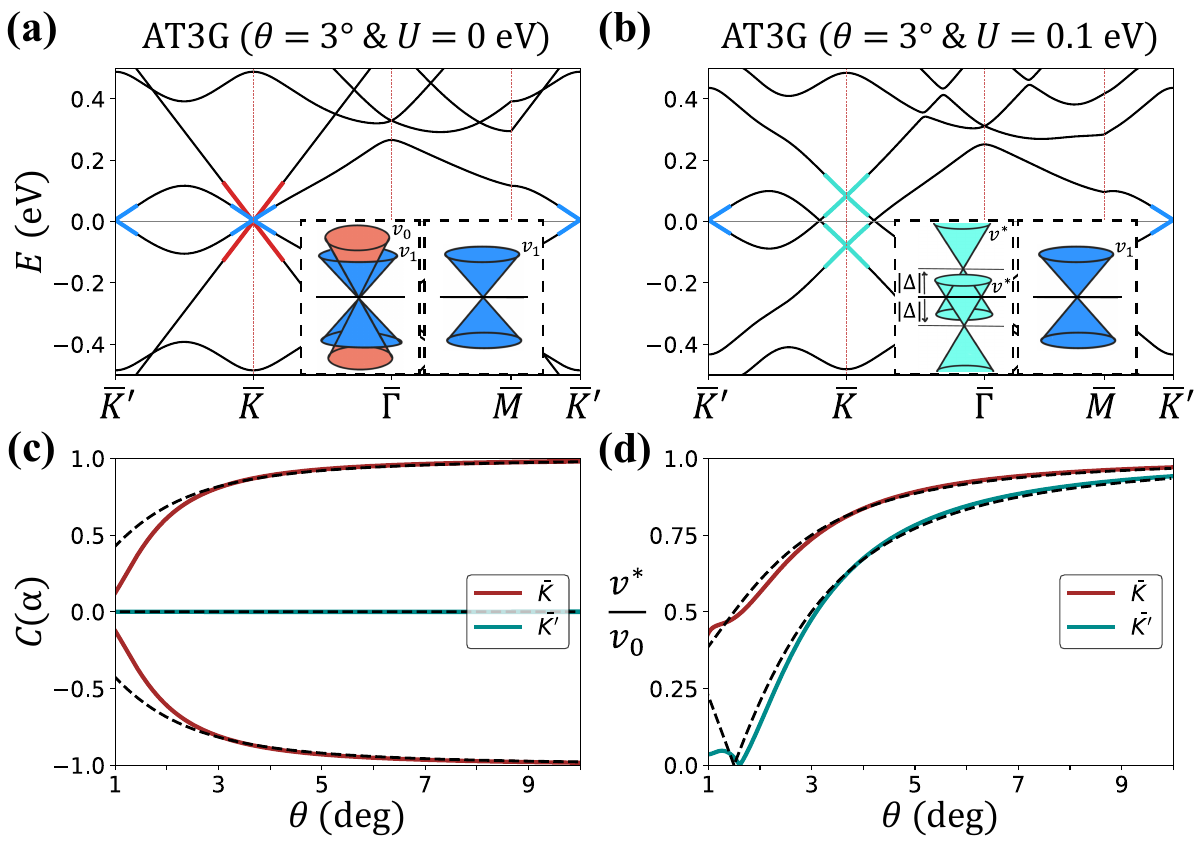}
\vspace{-5mm}
\caption{
Band structure of $N=3$ ATMG at $\theta=3^{\circ}$ for (a) $U=0$ and (b) $U=0.1$~eV. The left and right insets to (a) and (b) represent the schematic band structure near $\bar{K}$ and $\bar{K}'$. (c) $C(\alpha)$ and (d) $v^{\ast}/v_0$ as a function of twist angle $\theta$ for the full numerical calculations (solid line) and the analytical result from the rigid ($\omega$ = $\omega^{\prime}$) first shell model (dashed line).
} 
\label{fig:fig2}
\end{figure}

\begin{equation}
\label{eq:wave3K}
\Psi_{1, \lambda} = \frac{1}{\sqrt{2+24\alpha^{2}}}
\begin{pmatrix} 
a_{\lambda}\\
2b_{\lambda}\\
a_{\lambda}
\end{pmatrix},\;
\Psi_{2, \lambda} = \frac{1}{\sqrt{2}}
\begin{pmatrix} 
a_{\lambda}\\
0\\
-a_{\lambda}
\end{pmatrix}.\;
\end{equation}
At $\bar{K}$ in AT3G, the perturbation $\hat{V}$ in the first shell model is given by $\hat{V} = \rm{diag}$($-U\mathbb{I}_{2}, \boldsymbol{0}_{6}, U\mathbb{I}_{2}$). Then, in the basis of the wave functions in Eq.~(\ref{eq:wave3K}), we obtain the perturbation matrix $V_{\bar{K}}$ using $V_{11}=V_{22}=0$ and $V_{12}=V_{21}=-U/\sqrt{1+12\alpha^{2}}$. By diagonalizing $V_{\bar{K}}$, we obtain the effective Hamiltonian of biased AT3G near $\bar{K}$ as
\begin{equation}
H_{{\rm eff}, \bar{K}}^{(\pm)}=\pm \;C(\alpha)U+\hbar v^{*} (\boldsymbol{k} \cdot \boldsymbol{\sigma}),
\end{equation}
where $C(\alpha)=1/\sqrt{1+12\alpha^{2}}$ and $v^{*} = \left(v_{0}+v_{1}\right)/2$.
Comparing the left inset of Figs.~\ref{fig:fig2}(a) and~\ref{fig:fig2}(b), we can deduce that the two Dirac bands near $\bar{K}$ are hybridized equally and split by $2C(\alpha)U$ acquiring the average Fermi velocity $v^{*}$ from the unbiased values.

On the other hand, near $\bar{K'}$, only one Dirac cone with $v_{1}$ exists, whose wave function is
\begin{equation}
\label{eq:wave3Kp}
\Psi_{1, \lambda} = \frac{1}{\sqrt{1+12\alpha^{2}}}
\begin{pmatrix} 
b_{\lambda}\\
a_{\lambda}\\
b_{\lambda}
\end{pmatrix}.
\end{equation}
Then, the perturbation matrix $V_{\bar{K}'}$ vanishes and the Dirac cone at $\bar{K}'$ remains unaltered to leading order in $U$, resulting in the effective Hamiltonian
\begin{equation}
H_{{\rm eff}, \bar{K}'}=\hbar v_{1}(\boldsymbol{k} \cdot \boldsymbol{\sigma}).
\end{equation}
In Figs.~\ref{fig:fig2}(c) and~\ref{fig:fig2}(d), we illustrate the result of the leading-order energy splitting coefficient $C(\alpha)$ and the modified Fermi velocity $v^{*}$ obtained from the analytic model and numerical method as a function of twist angle.

\subsection{$N=4$}
\label{subSec:N=4}

\begin{figure}[t]
\includegraphics[width=1.0\linewidth]{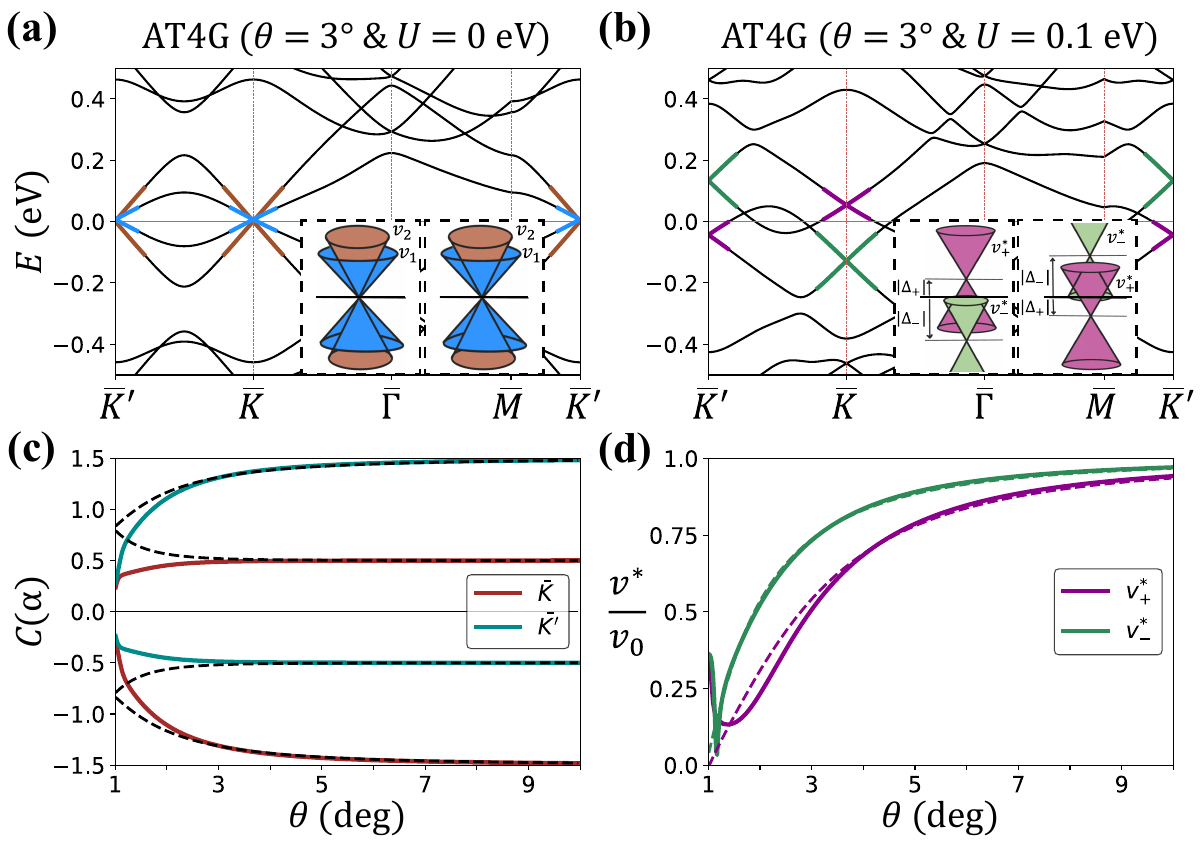}
\vspace{-5mm}
\caption{
Similar to panels (a)-(c) in Fig.~\ref{fig:fig2}, but for $N=4$ ATMG. If $U<0$, the energy shifts are reversed.
In panel (d), we show the two Fermi velocities $v_{\pm}^{*}$ as given in Eq.~(\ref{eq:vpm}) of the positively and negatively shifted Dirac cones illustrated in the inset to (b).
}
\label{fig:fig3}
\end{figure}

In the following, we derive the effective Hamiltonian of alternating twist tetralayer graphene (AT4G) at $\bar{K}$ and $\bar{K}'$. At $\bar{K}$, there are two Dirac cones with the velocities $v_{1}$ and $v_{2}$ as shown in Fig.~\ref{fig:fig3}(a), and the corresponding wave functions are given by
\begin{equation}
\Psi_{r, \lambda} = \frac{2}{\sqrt{5(1+6t_{r}^{2}\alpha^{2})}}
\begin{pmatrix} 
\sin{\theta_{r}}\cdot a_{\lambda}\\
\sin{2\theta_{r}}\cdot t_{r}b_{\lambda}\\
\sin{3\theta_{r}}\cdot a_{\lambda}\\
\sin{4\theta_{r}}\cdot t_{r}b_{\lambda}\\
\end{pmatrix}
\end{equation}
with $r=1, 2$. The perturbation $\hat{V}$ at $\bar{K}$ in this case is given by $\hat{V} = \rm{diag}$($-\frac{3U}{2}\mathbb{I}_{2}, -\frac{U}{2}\mathbb{I}_{6}, \frac{U}{2}\mathbb{I}_{2}, \frac{3U}{2}\mathbb{I}_{6}$). Since there are two Dirac cones at $\bar{K}$, the size of the perturbation matrix $V_{\bar{K}}$ would be $2\times2$, and its elements $V_{rr'}$ are expressed as
\begin{multline}
\label{eq:matrixelements}
V_{rr'} = \frac{4U}{N+1}\frac{1-6t_{r}t_{r'}\alpha^{2}}{\sqrt{(1+6t_{r}^{2}\alpha^{2})(1+6t_{r'}^{2}\alpha^{2})}}\\
\times\sum_{l=0}^{N/2}\left(2\ell-\frac{N-1}{2}\right)\sin{(2\ell+1)\theta_{r}}\sin{(2\ell+1)\theta_{r'}}\\
= \frac{2U}{5}\frac{1-6t_{r}t_{r'}\alpha^{2}}{\sqrt{(1+6t_{r}^{2}\alpha^{2})(1+6t_{r'}^{2}\alpha^{2})}}(-3\sin{\theta_{r}}\\
\times\sin{\theta_{r'}}+\sin{3\theta_{r}}\sin{3\theta_{r'}}+5\sin{5\theta_{r}}\sin{5\theta_{r'}}).
\end{multline}
By diagonalizing the matrix $V_{\bar{K}}$, we obtain the effective Hamiltonian of biased AT4G near $\bar{K}$ as
\begin{equation}
\label{eq:H_eff4K}
    H_{{\rm eff}, \bar{K}} = C_{\pm}(\alpha)U +\hbar v_{\pm}^{*}(\boldsymbol{k} \cdot \boldsymbol{\sigma}),
\end{equation}
where $C_{\pm}(\alpha)$ and $v_{\pm}^{*}$ corresponding to upward and downward shifted Dirac cones are given by
\begin{multline}
    C_{\pm}(\alpha) = \frac{1}{2(1+18\alpha^{2}+36\alpha^{4})}[-(1+12\alpha^{2}-36\alpha^{4})\\
    \pm\sqrt{(1+12\alpha^{2})(1+18\alpha^{2}+45\alpha^{4}+108\alpha^{6})}]
\end{multline}
and
\begin{equation}
    v^{*}_{\pm} = \frac{A_{\pm}^{2}v_{1}+B^{2}v_{2}}{A_{\pm}^{2}+B^{2}}.
    \label{eq:vpm}
\end{equation}
Here, $A_{\pm}$ and $B$ are unnormalized mixing coefficients of the two Dirac cones given, respectively, by
\begin{subequations}
\begin{eqnarray}
A_{\pm} &=& 1+15\alpha^{2}-36\alpha^{4} \\
&\pm&\sqrt{5(1+12\alpha^{2})(1+18\alpha^{2}+45\alpha^{4}+10    8\alpha^{6})}, \nonumber \\
B &=& -2(1+6\alpha^{2})\sqrt{1+18\alpha^{2}+36\alpha^{4}}.
\end{eqnarray}
\end{subequations}
From the above result, we find that the two Dirac cones with the velocities $v_{1}$ and $v_{2}$ are hybridized with the ratio of $A_{\pm}$ and $B$, and shifted by $C_{\pm}(\alpha)U$, as schematically illustrated in Fig.~\ref{fig:fig3}(b). 

On the other hand, near $\bar{K}'$, the wave functions for two Dirac cones with the velocity $v_{r}$ $(r=1, 2)$ are given by
\begin{equation}
\Psi_{r, \lambda} = \frac{2}{\sqrt{5(1+6t_{r}^{2}\alpha^{2})}}
\begin{pmatrix} 
\sin{\theta_{r}}\cdot t_{r}b_{\lambda}\\
\sin{2\theta_{r}}\cdot a_{\lambda}\\
\sin{3\theta_{r}}\cdot t_{r}b_{\lambda}\\
\sin{4\theta_{r}}\cdot a_{\lambda}\\
\end{pmatrix}.
\end{equation}
Since $\sin{\ell\theta_{r}}=(-1)^{r}\sin{(N+1-\ell)\theta_{r}}$, the wave function at $\bar{K}'$ can be obtained by reversing the components of the wave function at $\bar{K}$. Therefore, the effective Hamiltonian of biased AT4G near $\bar{K}'$ can be obtained by reversing the sign of the interlayer potential difference $U$ in Eq.~(\ref{eq:H_eff4K}) as
\begin{equation}
\label{eq:H_eff4Kp}
    H_{{\rm eff}, \bar{K}'} = -\;C_{\pm}(\alpha)U + \hbar v_{\pm}^{*}(\boldsymbol{k}\cdot\boldsymbol{\sigma}),
\end{equation}
where $C_{\pm}(\alpha)$ and $v_{\pm}^{*}$ are the same as those at $\bar{K}$. In detail, our model Hamiltonian [Eq.~(\ref{eq:modelH})] for $N=4$ has a combined symmetry expressed as 
\begin{equation}
\label{eq:H4sym}
(\hat{\Sigma}\hat{\mathcal{T}})H(\boldsymbol{k})(\hat{\Sigma}\hat{\mathcal{T}})^{-1} = -H(-\boldsymbol{k}),
\end{equation}
where
\begin{equation}
\label{eq:Sigma}
\hat{\Sigma}=
\begin{pmatrix}
0 & 0 & 0 & \sigma_{x}\\
0 & 0 & -\sigma_{x} & 0\\
0 & \sigma_{x} & 0 & 0\\
-\sigma_{x} & 0 & 0 & 0
\end{pmatrix}.
\end{equation}
We note that $\hat{\Sigma}$ changes only the valley index ($K \leftrightarrow K'$) keeping the same mBZ corner points ($\bar{K}\rightarrow\bar{K}$, $\bar{K}'\rightarrow\bar{K}'$)~\cite{Moon2013}, whereas the time-reversal operator $\hat{\mathcal{T}}$ changes both the valley index ($K \leftrightarrow K'$) and the mBZ corner points ($\bar{K}\leftrightarrow\bar{K}'$). This combined symmetry is preserved even in the presence of an interlayer potential difference, thus the effective Hamiltonians between $\bar{K}$ and $\bar{K}'$, which are respectively described in Eqs.~(\ref{eq:H_eff4K}) and~(\ref{eq:H_eff4Kp}), are also related as Eq.~(\ref{eq:H4sym}).

\subsection{Arbitrary $N$}
As the layer number $N$ is increased, the size of the perturbation matrix is also proportionally increased and it becomes progressively cumbersome to obtain analytically the effective Hamiltonian of ATMG for a large number of layers in the presence of an applied field even if we use the first shell model. Instead, here want to provide the general behavior patterns of the effective Hamiltonian of biased ATMG for arbitrary $N$. Tables~\ref{tab:H_eff_odd} and~\ref{tab:H_eff_even} show the summary of the effective Hamiltonian for $N=2$$-$$8$ ATMG in the presence of the interlayer potential difference.

\begin{table}[t]
\begin{ruledtabular}
\caption{Summary of the effective Hamiltonian of ATMG for odd numbers of layers $N=3, 5, 7$.}
\vspace{1mm}
\includegraphics[width=1.0\linewidth]{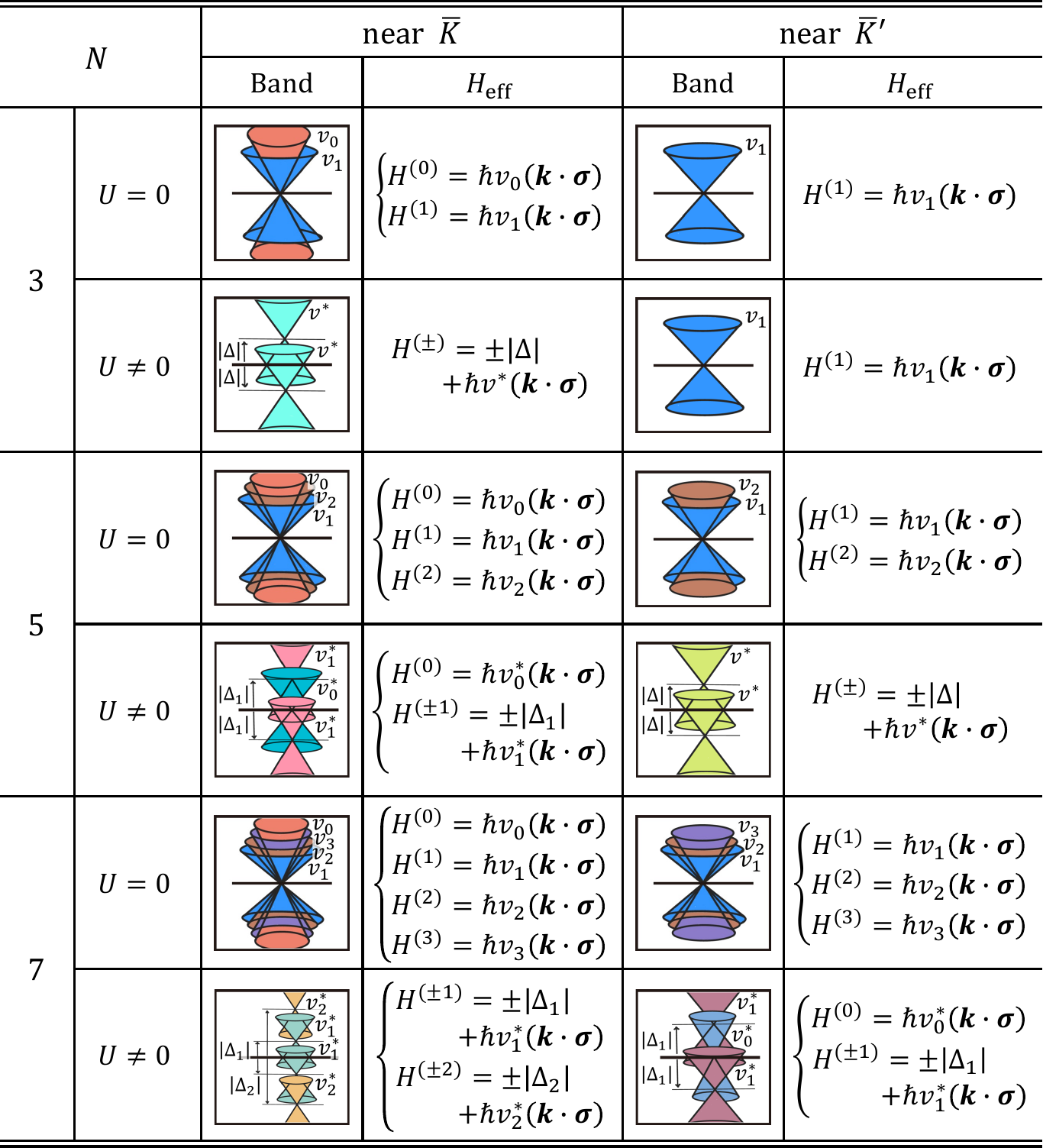}
\vspace{-5mm}
\label{tab:H_eff_odd}
\end{ruledtabular}
\end{table}

Firstly, for ATMG with an odd number of layers, there are $(N-1)/2$ TBG Dirac cones labeled by $v_{r}$ $\left[r=1,2,\ldots,(N-1)/2\right]$ near the two mBZ corners $\bar{K}$ and $\bar{K}'$, and one decomposed monolayer Dirac cone with $v_{0}$ at $\bar{K}$. Regardless of the layer number $N$ and mBZ symmetry points, the form of the perturbation matrix $V$ is solely determined by the number of Dirac cones at $\bar{K}$ or $\bar{K}'$, even though its elements depend on $N$.
One can thus find the same form of the effective Hamiltonian when the number of Dirac cones is the same, as seen in Table~\ref{tab:H_eff_odd}. In detail, if there are $m$ Dirac cones at $\bar{K}$ or $\bar{K}'$, we have $m/2$ pairs of Dirac-cones shifted by $\pm\Delta_{i}$ ($i=1,2,\ldots,m/2$) for even $m$, whereas we have $(m-1)/2$ pairs of Dirac cones plus one Dirac cone without energy shift for odd $m$. Each pair of Dirac cones has the same effective Fermi velocity.

\begin{table}[h]
\begin{ruledtabular}
\caption{Summary of the effective Hamiltonian of ATMG for even numbers of layers $N=2, 4, 6, 8$. Here, we assume $U>0$. For $U<0$, the energy shifts are reversed.}
\vspace{1mm}
\includegraphics[width=1.0\linewidth]{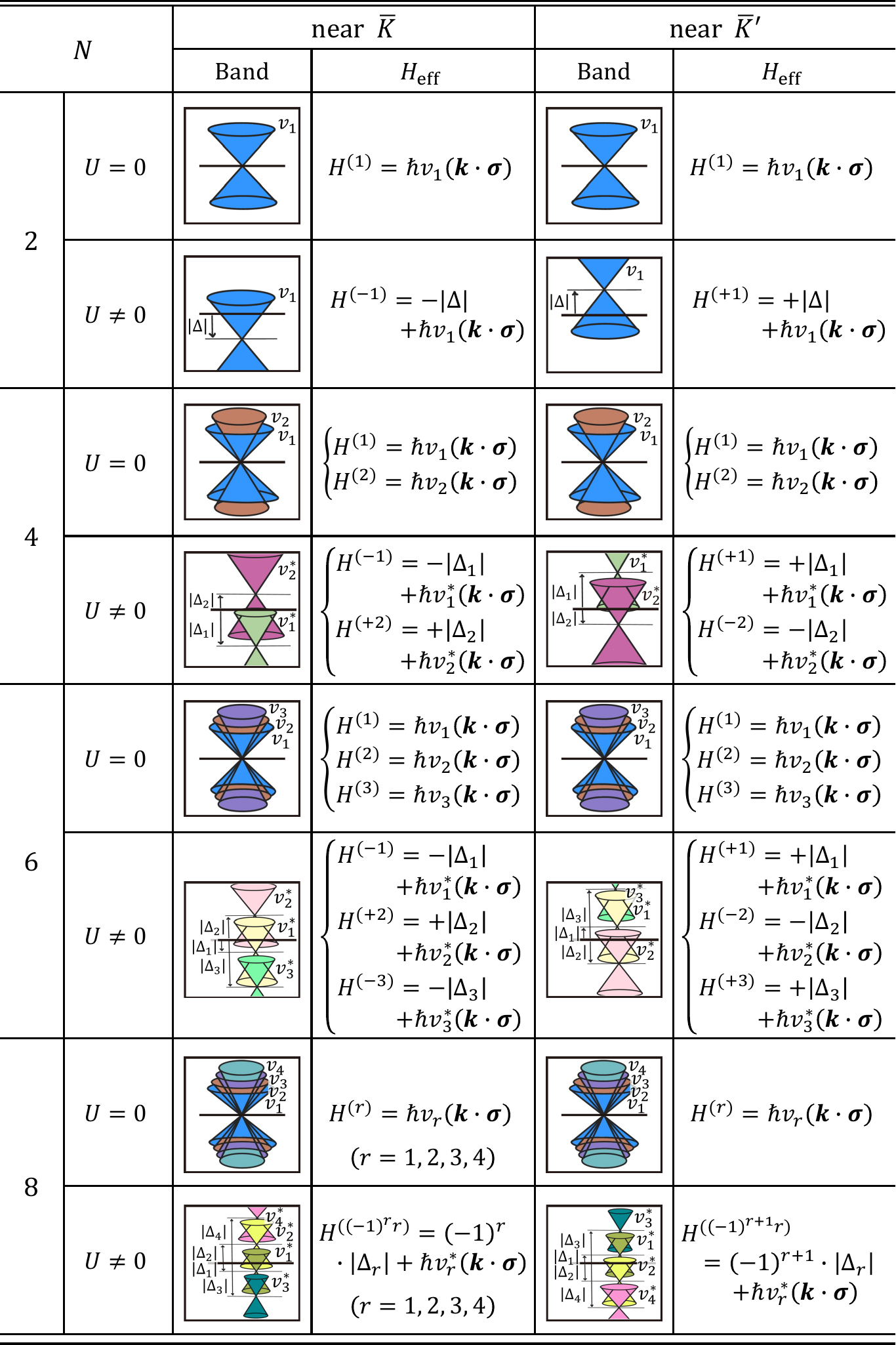}
\vspace{-5mm}
\label{tab:H_eff_even}
\end{ruledtabular}
\end{table}

For ATMG with an even number of layers, there are $N/2$ TBG Dirac cones labeled by $v_{r}$ $\left(r=1,2,\ldots, N/2\right)$ near $\bar{K}$ and $\bar{K}'$ when $U=0$. If an external field is applied ($U\neq0$), the Dirac cones are split with different energy shift $\Delta$ and Fermi velocity $v^{*}$. As already mentioned in Sec.~\ref{Sec:Sec2}.C, the effect of an applied electric field at $\bar{K}'$ ($\bar{K}$) can be effectively described by flipping its direction ($\hat{z}\rightarrow-\hat{z}$) at $\bar{K}$ ($\bar{K}'$).
Moreover, we can generalize Eq.~(\ref{eq:H4sym}) by expanding $\hat{\Sigma}$ symmetry in Eq.~(\ref{eq:Sigma})
by conveniently alternating $\sigma_{x}$ and $-\sigma_{x}$.
Similarly to the $N=4$ case, the combined $\hat{\Sigma}\hat{\mathcal{T}}$ symmetry is still preserved for ATMG with an even number of layers in the presence of an interlayer potential difference, relating the effective Hamiltonians between $\bar{K}$ and $\bar{K}'$ with flipped energy shifts, as seen in Table~\ref{tab:H_eff_even}.

\begin{figure}[htb]
\includegraphics[width=1.0\linewidth]{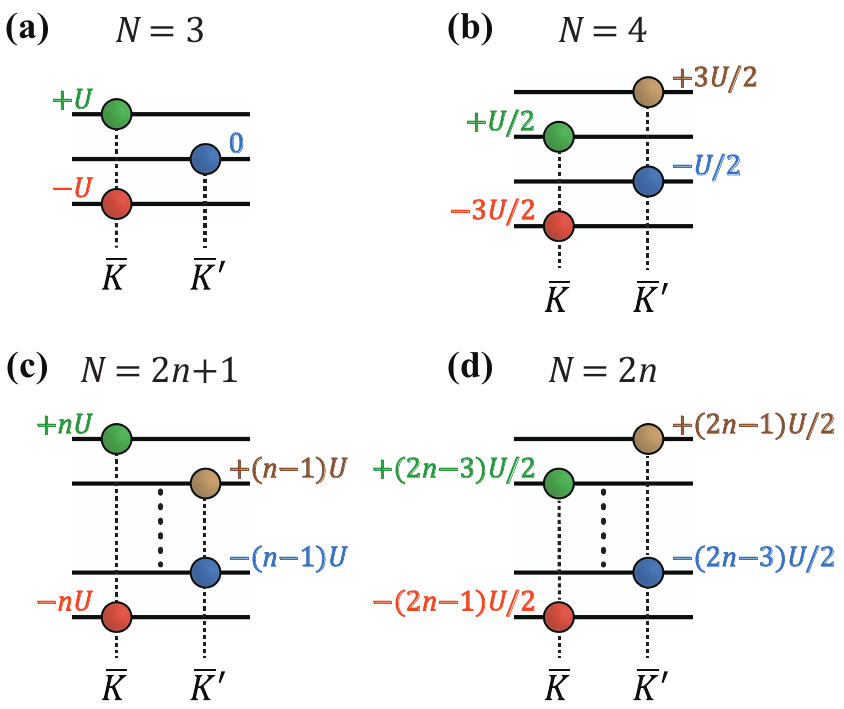}
\caption{Schematic picture of the energy shifts of Dirac cones at $\bar{K}$ and $\bar{K}'$ in biased ATMG in the asymptotic limit ($\alpha\rightarrow 0$).} 
\label{fig:fig4}
\end{figure}

Lastly, let us consider the effective Hamiltonian of biased ATMG in the asymptotic limit ($\alpha\rightarrow0$) where the twist angle $\theta$ becomes much larger than the first magic angle of ATMG. For the first shell model, $\lvert b_{\lambda} \rvert$ becomes proportional to $\alpha$, so only monolayer terms $a_{\lambda}$ of $\Psi_{r,\lambda}$ survive in this limit. 
Thus, the energy splitting coefficient $C(\alpha)$ of ATMG with arbitrary $N$ at $\bar{K}$ ($\bar{K}'$) can be obtained as odd (even) layer components of $\hat{V}$, as schematically shown in Fig.~\ref{fig:fig4}.

\begin{figure}[htb]
\includegraphics[width=1.0\linewidth]{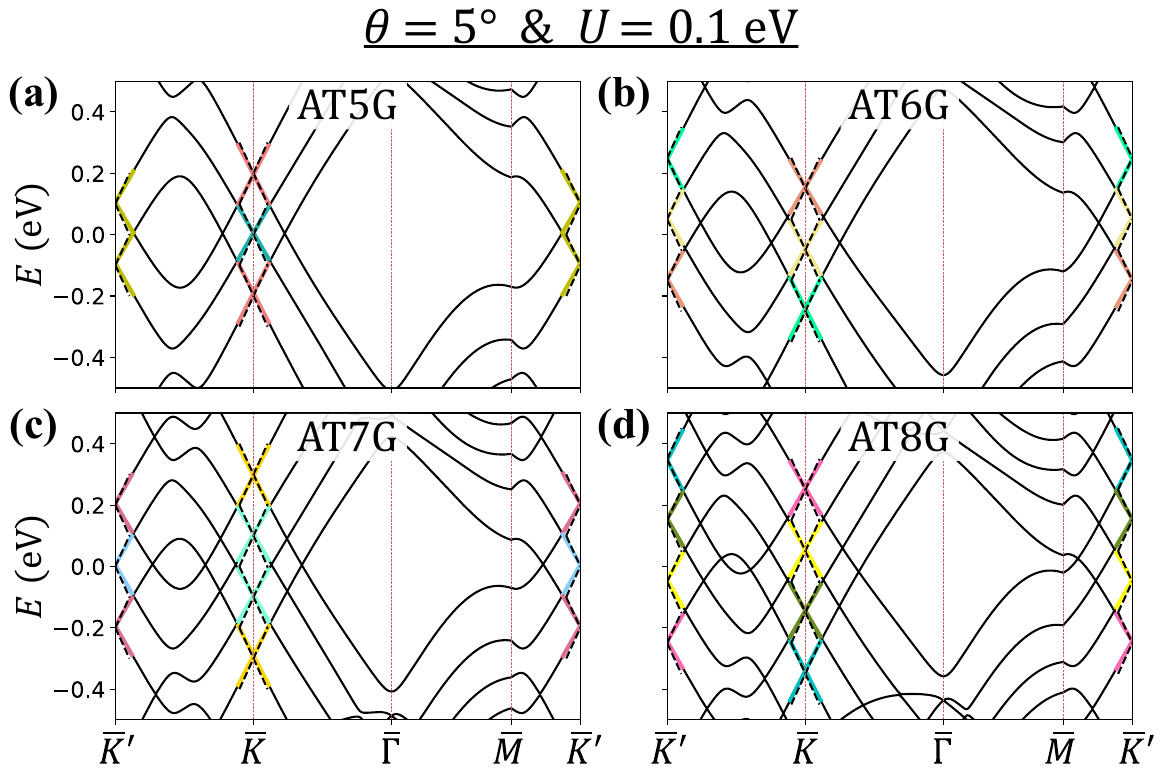}
\vspace{-5mm}
\caption{Band structure of $N=5-8$ ATMG at $\theta=5^{\circ}$ with $U=0.1$~eV. 
Solid and dashed lines represent the numerical calculations and the analytical result obtained in the $\alpha\rightarrow0$ limit, respectively.
}
\label{fig:fig5}
\end{figure}

On the other hand, the modified Fermi velocity $v^{*}$ converges to $v_{0}$ as $\alpha\rightarrow0$, since all eigenstates of biased ATMG in this limit have just a single monolayer term, giving the monolayer graphene Dirac cone with the velocity $v_{0}$.
Therefore, the effective Hamiltonian would be described by a set of monolayer graphene Hamiltonian with the energy shift described by $C(\alpha\rightarrow0)U$, which can be obtained by the pattern presented in Fig.~\ref{fig:fig4}.
Figure~\ref{fig:fig5} shows the band structure of $N=5-8$ ATMG in the presence of the interlayer potential difference $U=0.1$~eV at $\theta=5^{\circ}$ along with the analytical result obtained in the asymptotic limit, which agrees closely with the full numerical result except for small deviations in the Fermi velocity of the Dirac cones.

\section{Optical conductivity}
\label{Sec:Sec3}

The Kubo formula for the optical conductivity in the non-interacting and clean limit is given by~\cite{Mahan2000}
\begin{eqnarray}
\label{eq:conductivity}
\sigma_{ij}(\omega)
&=&- \frac{ie^2}{\hbar} \sum_{s,s'} \int \frac{d^2 k}{(2\pi)^2} \frac{f_{s, \bm{k}}-f_{s',\bm{k}}}{\varepsilon_{s,\bm{k}}-\varepsilon_{s',\bm{k}}} \nonumber \\
&\times&\frac{M^{ss'}_i(\bm k)M^{s's}_j(\bm k)}{\hbar\omega+\varepsilon_{s,\bm{k}}-\varepsilon_{s',\bm{k}}+i0^{+}},
\end{eqnarray}
where $i,j=x,y$, $f_{s,\bm{k}}=1/[1+e^{(\varepsilon_{s,\bm{k}}-\mu)/k_{\rm B}T}]$ is the Fermi distribution function for the band index $s$ and wave vector $\bm{k}$, $\mu$ is the chemical potential and $M^{ss'}_i(\bm k)=\langle{s,\bm{k}}|\hbar\hat{v}_i |{s',\bm{k}}\rangle$ with the velocity operator $\hat{v}_i$ obtained from the relation $\hat{v}_i=\frac{1}{\hbar}\frac{\partial \hat{H}}{\partial  k_i}$.

\begin{figure}[b]
\includegraphics[width=1.0\linewidth]{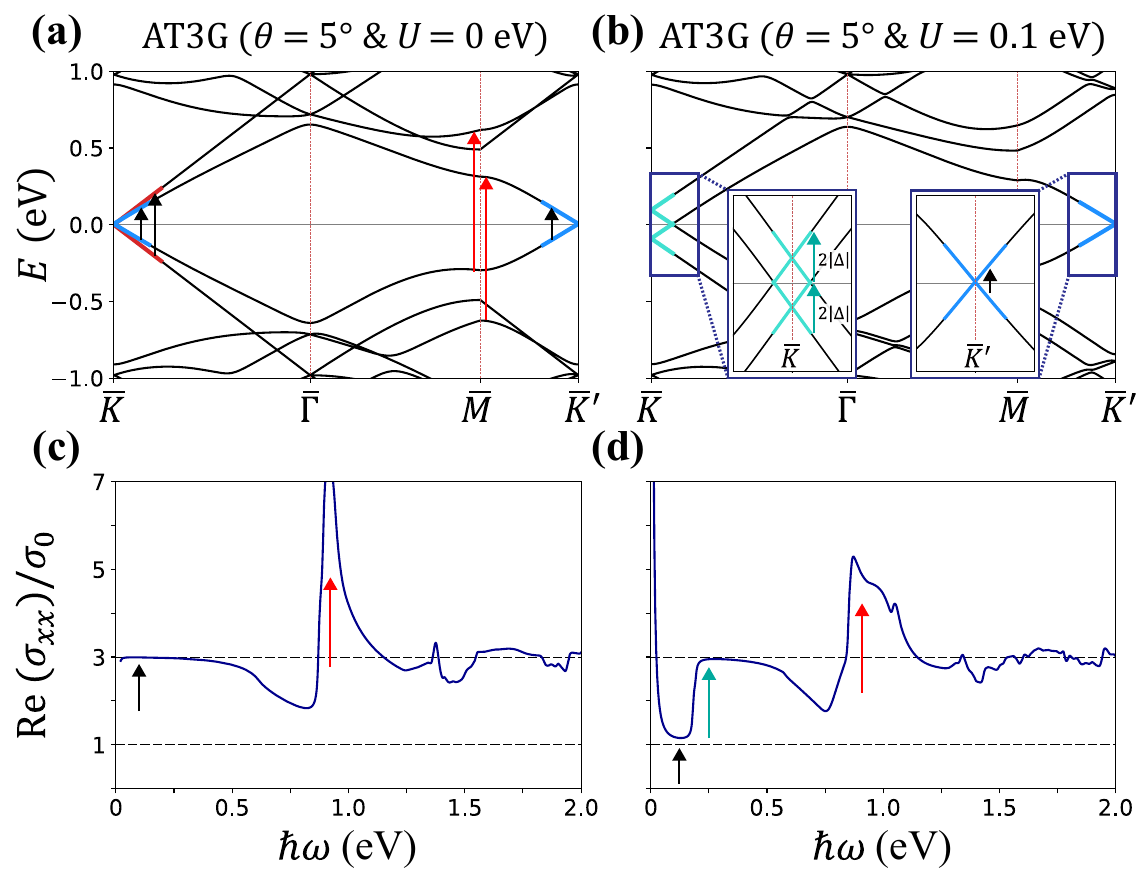}
\vspace{-5mm}
\caption{
Band structure and the longitudinal conductivity of $N=3$ ATMG at $\theta=5^{\circ}$ for (a), (c) $U=0$ and (b), (d) $U=0.1$~eV. The insets to (b) show an enlarged view of the band structure near $\bar{K}$ and $\bar{K}'$. The arrows in the band structure indicate interband transitions corresponding to peaks in the conductivity. In (d), a Drude peak appears at low frequencies due to intraband contributions.
} 
\label{fig:fig6}
\end{figure}

\begin{figure}[t]
\includegraphics[width=1.0\linewidth]{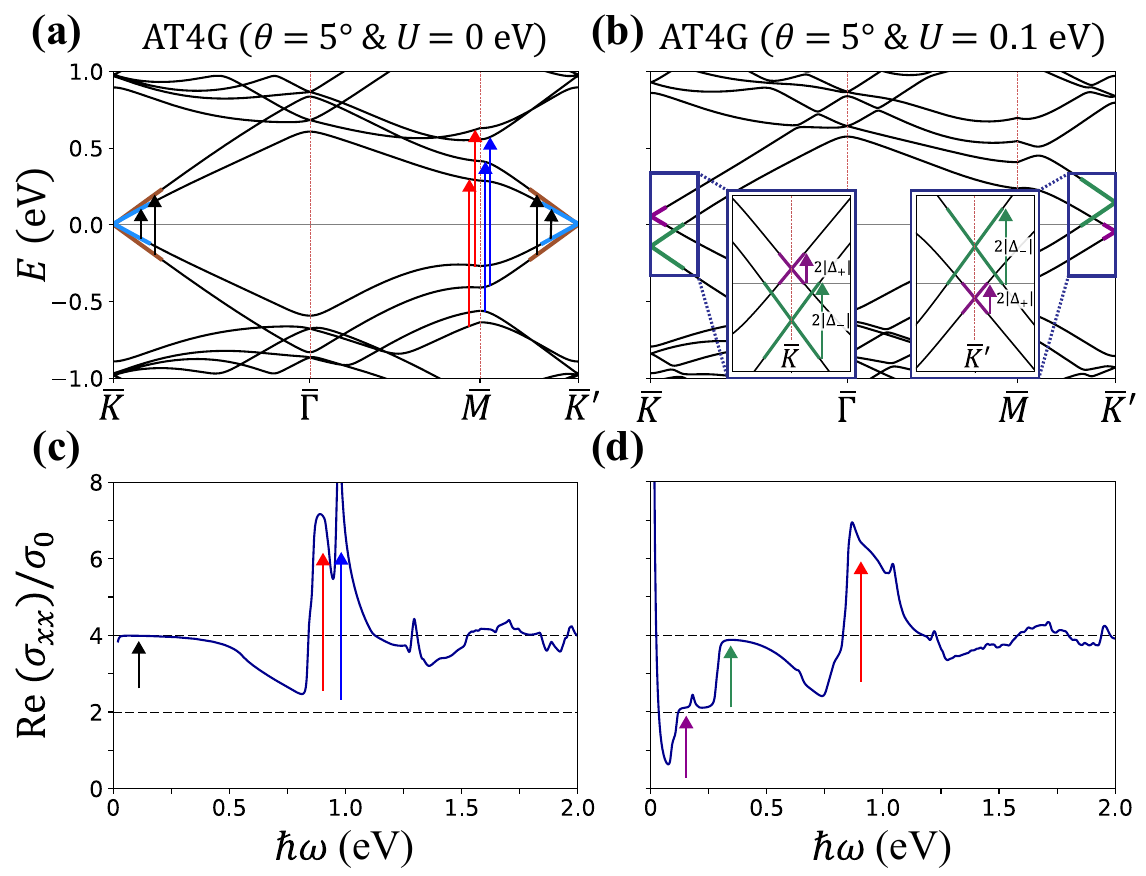}
\vspace{-5mm}
\caption{
Same as Fig.~\ref{fig:fig6} for $N=4$ ATMG.
} 
\label{fig:fig7}
\end{figure}

In the following, we consider the real part of the longitudinal optical conductivity for $\mu=0$ at zero temperature with a finite broadening term $\eta = 5$~meV replacing the $0^{+}$ term in Eq.~(\ref{eq:conductivity}) for numerical calculations.
We plot the optical conductivities of AT3G and AT4G for the continuum model [Eq.~(\ref{eq:modelH})] with and without the interlayer potential difference $U$ at the twist angle $\theta=5^{\circ}$ in Figs.~\ref{fig:fig6} and~\ref{fig:fig7}, respectively.


In the absence of the interlayer potential difference, the longitudinal conductivity converges to $N\sigma_{0}$ for both low- and high-frequency limits, as shown in Figs.~\ref{fig:fig6}(c) and~\ref{fig:fig7}(c). Here, $\sigma_{0}=g_{\rm sv}e^{2}/16\hbar$ is the optical conductivity of charge-neutral monolayer graphene with the spin-valley degeneracy factor $g_{\rm sv}=4$. The low-frequency conductivity originates from transitions within $N$ hybridized Dirac nodes located at $\bar{K}$ and $\bar{K}'$, whereas at high frequencies the interlayer coupling becomes negligible thus the conductivity approaches that of $N$ decoupled monolayer graphene sheets.
At intermediate frequencies, a dominant peak appears around $\hbar\omega \sim 0.9$~eV for $\theta = 5^{\circ}$ arising from interband transitions between states near the saddle point $\bar{M}$, as indicated by the red arrows. The frequency where the dominant peak appears depends on the twist angle $\theta$ but weakly depends on $N$ or $U$.

In the presence of the interlayer potential difference, the conductivity shows a step-like feature at low frequencies, as shown in Figs.~\ref{fig:fig6}(d) and~\ref{fig:fig7}(d).
For biased AT3G, two Dirac nodes at $\bar{K}$ are split by $2\Delta$, thus, interband transitions are forbidden in the low-frequency limit, while interband transitions are allowed in the unaltered Dirac cone at $\bar{K}'$, giving $\sigma_{0}$.
For biased AT4G, two Dirac nodes at $\bar{K}$ and another two Dirac nodes at $\bar{K}'$ are shifted by $\Delta_{\pm}$, thus, interband transitions are forbidden in the low-frequency limit, and the optical conductivity vanishes.
As the frequency increases, the optical conductivity increases by $2\sigma_{0}$ when $\hbar\omega\sim 2\lvert\Delta\rvert$ and $2\lvert\Delta_{\pm}\rvert$ for biased AT3G and AT4G, respectively, eventually approaching $N\sigma_{0}$.
This feature is very analogous to the optical conductivity of AA-stacked multilayer graphene, where the optical conductivity increases in steps of $2\sigma_{0}$ toward $N\sigma_{0}$ when interband transitions occur within the same Dirac cones~\cite{min2009,Tabert2012}.
Unlike AA-stacked multilayer graphene, however, the velocity changes away from $\bar{K}$ or $\bar{K}'$, so the transition energy deviates from $2\lvert\Delta\rvert$ and $2\lvert\Delta_{\pm}\rvert$, especially at small twist angles.
As the twist angle decreases, interband transitions arising from the $\bar{M}$ and $\bar{\Gamma}$ points occur at lower energies and eventually mix with the interband transitions arising from the $\bar{K}$ and $\bar{K}'$ points, blurring the step-like features explained above.
The evolution of the optical conductivity with decreasing twist angle will be discussed in Appendix~\ref{Sec:Magic3-4}.

\section{Discussion}
\label{Sec:Sec4}

The analytical forms of the effective Hamiltonian for biased ATMG near $\bar{K}$ or $\bar{K}'$ were obtained using the first shell model of the moir\'{e} $\bm{G}$ vectors, which is valid within the radius about $k_{c}\sim U/\hbar v_{0}$ in the $\bm{k}$ space beyond which two shifted Dirac cones by $U$ cross each other.
When the twist angle becomes smaller than the typical values of the first magic angle $\theta_{\rm M}^{(N)}\leq\theta_{\rm M}^{(\infty)}\approx2.2^{\circ}$~\cite{Khalaf2019}, our first shell model, which employs the nearest-neighbor truncation, is generally insufficient for accurately capturing the bands of an enlarged moir\'{e} superlattice of ATMG, resulting in the discrepancy between analytical and numerical results.
Nevertheless, the analytical results obtained from our perturbation approach agree well with the full numerical calculations for twist angles $\theta\gtrsim 2.2^{\circ}$ where the interlayer coupling is weaker.

In summary, we have studied the effect of a perpendicular electric field on ATMG, focusing on the effects of an interlayer potential difference in altering the low-energy band structure and therefore the optical absorption spectrum, which can be used as a distinguishing experimental signature.
Firstly, we analytically derived the low-energy effective Hamiltonian and its energy spectrum near the two moir\'{e} Dirac points $\bar{K}$ and $\bar{K}'$ up to pentalayer by using first-order degenerate-state perturbation theory, treating the asymmetric interlayer potential difference as a perturbation.
Then, we presented general rules for constructing the effective Hamiltonian of biased ATMG with an arbitrary number of layers.
Lastly, we investigated the optical absorption spectrum of ATMG with and without an interlayer potential difference.
We found that the longitudinal conductivity of biased ATMG showed a step-like feature arising from the splitting of Dirac nodes by the applied electric field, which is reminiscent of the optical conductivity features of AA-stacked multilayer graphene.

\acknowledgments
This work was supported by the National Research Foundation of Korea (NRF) grant funded by the Korea government (MSIT) (Grant No. 2018R1A2B6007837 and No. 2023R1A2C1005996), the Creative-Pioneering Researchers Program through Seoul National University (SNU), and the Center for Theoretical Physics. 
J. S. was supported by Korea NRF (Grant No. 2021R1A6A3A01087281), and J. J. acknowledges support from Korea NRF (Grant No. 2020R1A2C3009142). 

\appendix
\section{Derivation of the effective Hamiltonian for $N=5$ ATMG}
\label{Sec:N=5}

\begin{figure}[t]
\includegraphics[width=1.0\linewidth]{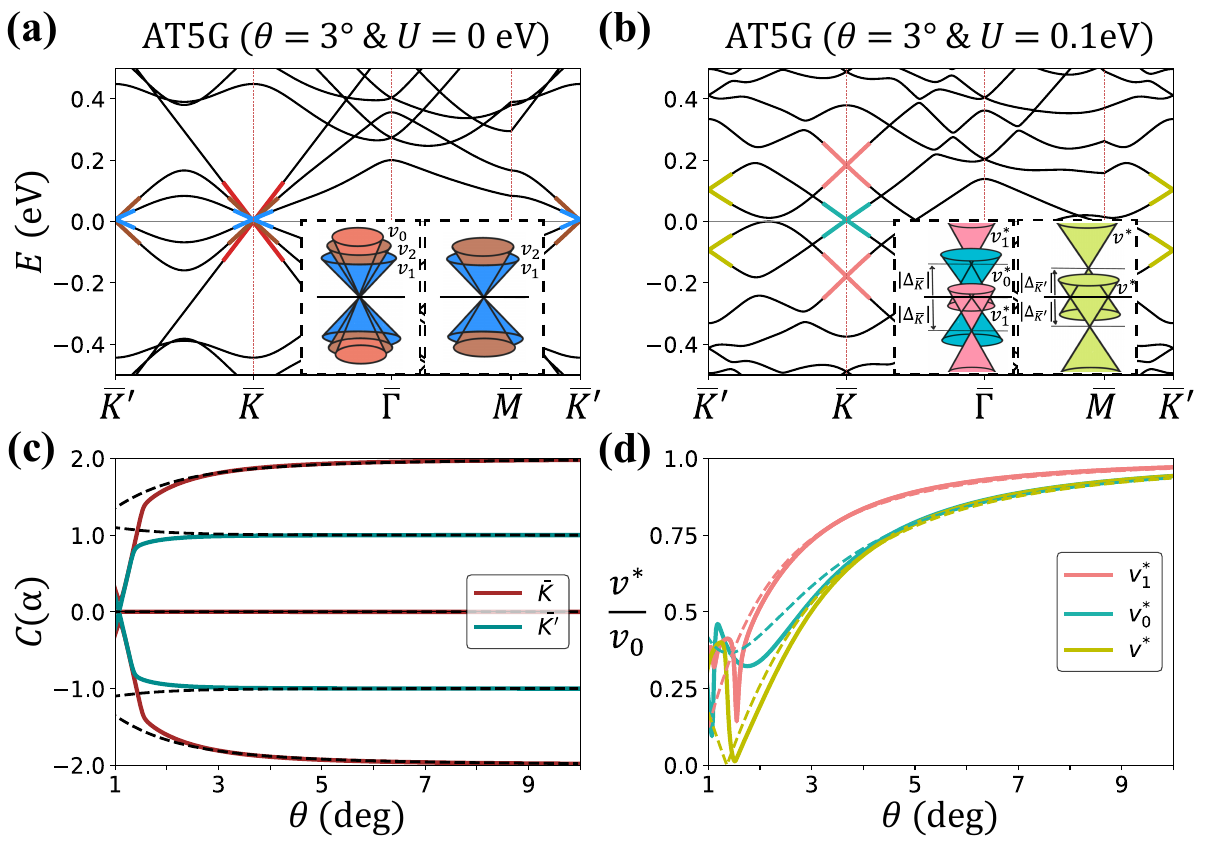}
\vspace{-5mm}
\caption{
Similar to panels (a)-(c) in Fig.~\ref{fig:fig2}, but for $N=5$ ATMG. In panel (d), we show the modified Fermi velocities $v_{0}^{*}$ and $v_{1}^{*}$ given in Eq.~(\ref{eq:v01}) at $\bar{K}$ and $v^{*}=(v_1+v_2)/2$ at $\bar{K}'$ as illustrated in the inset to (b).
}
\label{fig:fig8}
\end{figure}

In this Appendix, we derive the effective Hamiltonian of alternating twist pentalayer graphene (AT5G) at the two moir\'{e} Dirac points $\bar{K}$ and $\bar{K}'$. Three Dirac cones labeled by $v_{0}$, $v_{1}$, and $v_{2}$ exist near $\bar{K}$ as shown in Fig.~\ref{fig:fig8}(a), thus the size of the perturbation matrix $V_{\bar{K}}$ is $3\times3$. Using Eq.~(\ref{eq:Psi_r}), we obtain the following normalized wave functions $\Psi_{r, \lambda}$ near $\bar{K}$ in our first shell model:
\begin{subequations}
\label{eq:wave5K}
    \begin{equation}
    \Psi_{r, \lambda} = \frac{2}{\sqrt{6(1+6t_{r}^{2}\alpha^{2})}}
    \begin{pmatrix} 
    \sin{\theta_{r}}\cdot a_{\lambda}\\
    \sin{2\theta_{r}}\cdot t_{r}b_{\lambda}\\
    \sin{3\theta_{r}}\cdot a_{\lambda}\\
    \sin{4\theta_{r}}\cdot t_{r}b_{\lambda}\\
    \sin{5\theta_{r}}\cdot a_{\lambda}
    \end{pmatrix},
    \end{equation}
    \begin{equation}
    \Psi_{3, \lambda} = \frac{1}{\sqrt{3}}
    \begin{pmatrix} 
    a_{\lambda}\\
    0\\
    -a_{\lambda}\\
    0\\
    a_{\lambda}\\
    \end{pmatrix}
    \end{equation}
\end{subequations}
with $r=1, 2$. At $\bar{K}$ in AT5G, the perturbation $\hat{V}$ is given by $\hat{V} = \rm{diag}$($-2U\mathbb{I}_{2\times2}$, $-U\mathbb{I}_{6\times6}$, $\boldsymbol{0}_{2\times2}$, $U\mathbb{I}_{6\times6}$, $2U\mathbb{I}_{2\times2}$). Following the same procedure described in Secs.~\ref{Sec:Sec2}.B and~\ref{Sec:Sec2}.C, we obtain the perturbation matrix $V_{\bar{K}}$ with the elements of $V_{11}=V_{22}=V_{33}=V_{13}=V_{31}=0$, $V_{12}=V_{21}=-2U(1+9\alpha^{2})/\sqrt{3(1+6\alpha^{2})(1+12\alpha^{2})}$, and $V_{23}=V_{32}=-4U/\sqrt{6(1+6\alpha^{2})}$ in the basis of the wave functions in Eq.~(\ref{eq:wave5K}). Therefore, we obtain the effective Hamiltonian of biased AT5G near $\bar{K}$ by diagonalizing $V_{\bar{K}}$ as
\vspace{-0.5mm}
\begin{subequations}
    \begin{eqnarray}
    H^{(0)}_{{\rm eff}, \bar{K}} &=& \hbar v_{0}^{*}(\boldsymbol{k} \cdot \boldsymbol{\sigma}), \\
    H^{(\pm1)}_{{\rm eff}, \bar{K}} &=& \pm C_{\bar{K}}(\alpha)U+\hbar v_{1}^{*}(\boldsymbol{k}\cdot\boldsymbol{\sigma}),
    \end{eqnarray}
\end{subequations}
\vspace{-0.5mm}
where
\vspace{-0.5mm}
\begin{equation}
C_{\bar{K}}(\alpha) = 2\sqrt{\frac{1+18\alpha^{2}+27\alpha^{4}}{(1+6\alpha^{2})(1+18\alpha^{2})}},
\vspace{-0.5mm}
\end{equation}
and
\begin{equation}
\label{eq:v01}
v_{0}^{*} = \frac{A^{2}v_{0}+B^{2}v_{1}}{A^{2}+B^{2}}, \;\; v_{1}^{*} = \frac{B^{2}v_{0}+A^{2}v_{1}}{2(A^{2}+B^{2})}+\frac{v_{2}}{2}.
\vspace{-1mm}
\end{equation}
Here, $A=1+9\alpha^{2}$ and $B=\sqrt{2(1+18\alpha^{2})}$ are unnormalized mixing coefficients of the Dirac cones.

On the other hand, near $\bar{K}'$, there are two Dirac cones with the velocities $v_{1}$ and $v_{2}$ as shown in Fig.~\ref{fig:fig8}(a), and the corresponding wave functions are given by
\begin{equation}
\Psi_{r, \lambda} = \frac{2}{\sqrt{6(1+6t_{r}^{2}\alpha^{2})}}
\begin{pmatrix} 
\sin{\theta_{r}}\cdot t_{r}b_{\lambda}\\
\sin{2\theta_{r}}\cdot a_{\lambda}\\
\sin{3\theta_{r}}\cdot t_{r}b_{\lambda}\\
\sin{4\theta_{r}}\cdot a_{\lambda}\\
\sin{5\theta_{r}}\cdot t_{r}b_{\lambda}
\end{pmatrix}
\end{equation}
with $r=1, 2$. Then, the size of perturbation matrix $V_{\bar{K}'}$ would be $2\times2$ with the elements of $V_{11}=V_{22}=0$ and $V_{12}=V_{21}=-U(1+12\alpha^{2})/\sqrt{(1+6\alpha^{2})(1+18\alpha^{2})}$. By diagonalizing $V_{\bar{K}'}$, we obtain the effective Hamiltonian of biased AT5G near $\bar{K}'$ as
\begin{equation}
H^{(\pm)}_{{\rm eff}, \bar{K}'} = \pm\;C_{\bar{K}'}(\alpha)U + \hbar v^{*}(\boldsymbol{k}\cdot\boldsymbol{\sigma}),
\end{equation}
where
\begin{equation}
C_{\bar{K}'} = \frac{1+12\alpha^{2}}{\sqrt{(1+6\alpha^{2})(1+18\alpha^{2})}}
\end{equation}

\begin{figure}[t]
\includegraphics[width=1.0\linewidth]{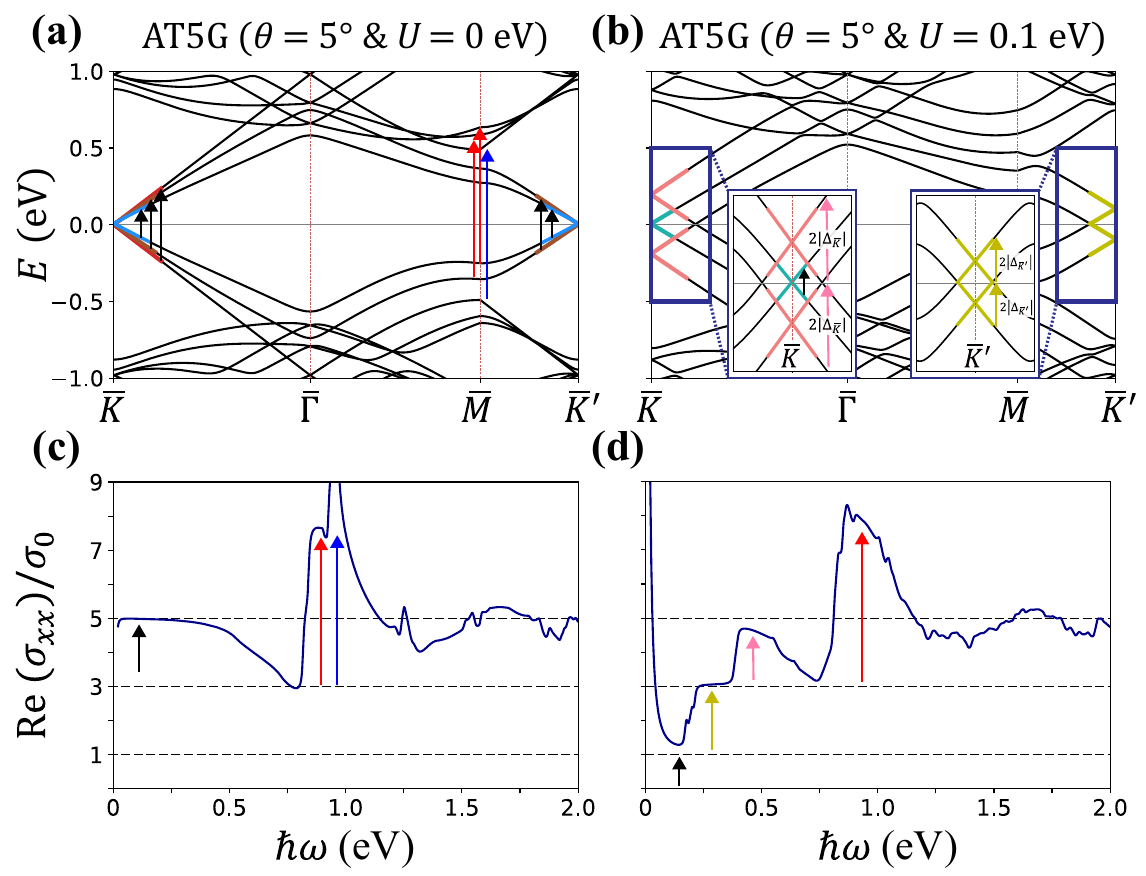}
\vspace{-5mm}
\caption{
Same as Fig.~\ref{fig:fig6} for $N=5$ ATMG.
}
\label{fig:fig9}
\end{figure}
\noindent
and $v^{*} = (v_{1}+v_{2})/2$. We here notice that the effective Hamiltonian of biased AT5G near $\bar{K}'$ has a similar form of one at biased AT3G near $\bar{K}$ since the number of Dirac cones are the same. In both cases, two Dirac cones are shifted by $\pm C(\alpha)U$ and hybridized equally, so that the equal Fermi velocity $v^{*}$ is assigned to be an average of unbiased ones. The only difference between them is the value of off-diagonal matrix elements, determining the energy shift of Dirac cones.

\section{Optical conductivity of $N=5$ ATMG}
\label{Optical:N=5}

In this Appendix, we present the real part of the longitudinal conductivity of AT5G with and without the interlayer potential difference $U$. Figure~\ref{fig:fig9} illustrates the longitudinal conductivity of AT5G for $U=0$ and $U=0.1$~eV at the twist angle $\theta = 5^{\circ}$, respectively.

\begin{figure}[tbh!]
\includegraphics[width=1.0\linewidth]{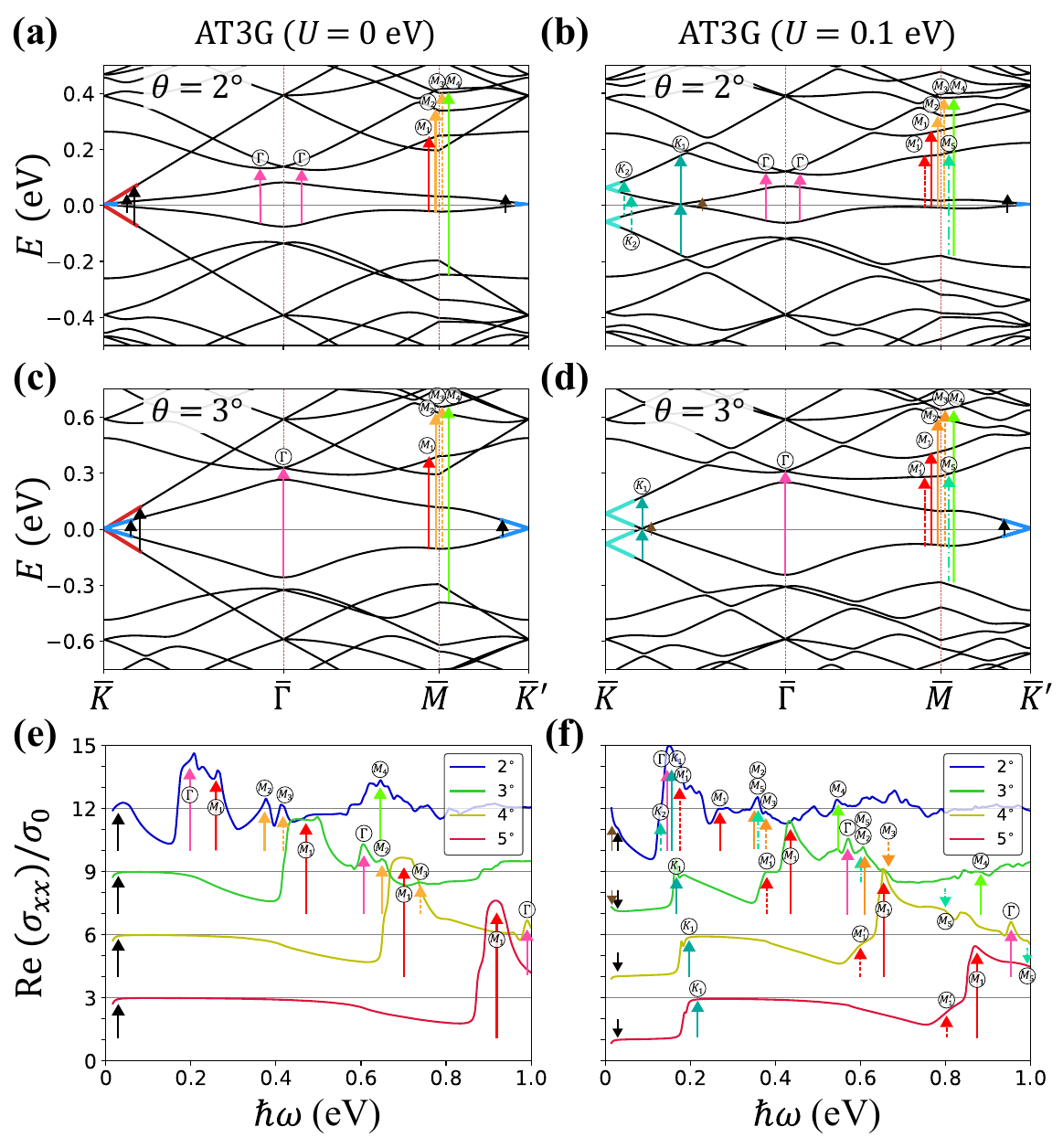}
\vspace{-5mm}
\caption{
Band structure of $N=3$ ATMG at $\theta=2^{\circ}$ and $\theta=3^{\circ}$ for (a), (c) $U=0$ and (b), (d) $U=0.1$~eV, respectively, and the evolution of the longitudinal optical conductivity with decreasing twist angle for (e) $U=0$ and (f) $U=0.1$~eV. The arrows in the band structure indicate interband transitions corresponding to peaks in the conductivity.
}
\label{fig:fig10}
\end{figure}

In the absence of the interlayer potential difference, the longitudinal conductivity converges to $5\sigma_{0}$ for both low- and high-frequency regions, as shown in Fig.~\ref{fig:fig9}(c).
On the other hand, in the presence of the interlayer potential difference, the conductivity shows a step-like feature at low frequencies, as shown in Fig.~\ref{fig:fig9}(d).
Specifically, the conductivity starts with $\sigma_{0}$ from the unshifted Dirac cone at $\bar{K}$ then increases toward $5\sigma_{0}$ in steps of $2\sigma_{0}$ when $\hbar\omega\sim 2\lvert\Delta_{\bar{K}}\rvert$ and $2\lvert\Delta_{\bar{K}'}\rvert$, respectively, at which the forbidden interband transitions due to the splitting of Dirac nodes by the applied electric field can occur.
The conductivity jump at $\hbar \omega\sim2\lvert\Delta_{\bar{K}}\rvert$, however, is only $1.5\sigma_{0}$ less than $2\sigma_{0}$, approaching the conductivity value for $U=0$ at that frequency.
This mismatch is due to the continuous decrease of the optical conductivity as the frequency increases because interband transitions are no longer described by those between the Dirac nodes near $\bar{K}$ and $\bar{K}'$.
When smaller $U$ or larger $\theta$ is used, one may see more clearly the conductivity increase in steps of $2\sigma_{0}$.
At intermediate frequencies, a dominant peak arises from interband transitions near $\bar{M}$, as indicated by the red arrows.

\section{Evolution of the optical conductivity with decreasing-twist angle}
\label{Sec:Magic3-4}

In the following, we consider the evolution of the longitudinal optical conductivities with decreasing twist angle for $\mu=0$ at zero temperature with a smaller broadening term $\eta = 3$~meV compared to that used in Figs.~\ref{fig:fig6} and~\ref{fig:fig7} to capture the low-frequency behavior more accurately.

\begin{figure}[t!]
\includegraphics[width=1.0\linewidth]{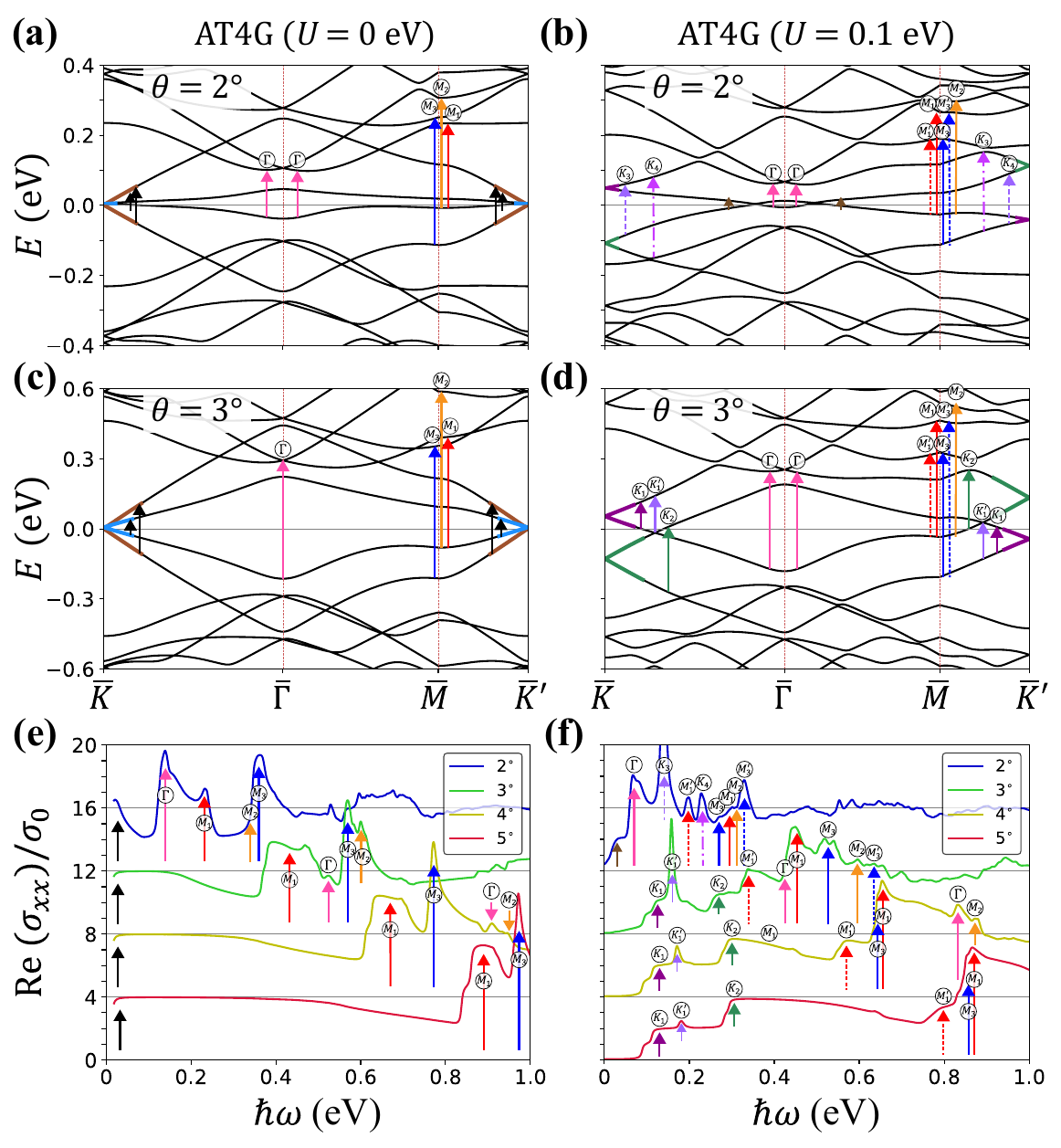}
\vspace{-5mm}
\caption{
Same as Fig.~\ref{fig:fig10} for $N=4$ ATMG.
} 
\label{fig:fig11}
\end{figure}

We plot the optical conductivities of AT3G and AT4G with and without the interlayer potential difference $U$ at

\newpage
\noindent
various twist angles $\theta=2^{\circ}$$-$$5^{\circ}$ in Figs.~\ref{fig:fig10} and~\ref{fig:fig11}, respectively.
Notice that in this section we only consider interband transitions, ignoring the Drude peak arising from intraband transitions.

When the interlayer potential difference is absent, the longitudinal conductivities converge to $N\sigma_{0}$ in the low-frequency limit but drop more quickly as the twist angle decreases due to the decrease in the bandwidth, as shown in Figs.~\ref{fig:fig10}(e) and~\ref{fig:fig11}(e).
Furthermore, interband transitions arising from the $\bar{M}$ and $\bar{\Gamma}$ points
[see Figs.~\ref{fig:fig10}(a), \ref{fig:fig10}(c), \ref{fig:fig11}(a), and \ref{fig:fig11}(c)],
which were regarded as high-energy transitions in Sec.~\ref{Sec:Sec3}, occur at lower energies, and the corresponding peaks move toward the low-frequency region as the twist angle decreases.

When the interlayer potential difference is present, the step-like feature discussed in Sec.~\ref{Sec:Sec3} can still be observed due to the interband transitions within the same Dirac cones, as shown in Figs.~\ref{fig:fig10}(f) and ~\ref{fig:fig11}(f).
However, as the twist angle decreases, interband transitions arising from the $\bar{M}$ and $\bar{\Gamma}$ points occur at lower energies and eventually mix with the interband transitions arising from the $\bar{K}$ and $\bar{K}'$ points, blurring the step-like features.
Furthermore, unlike AA-stacked multilayer graphene, the velocity changes away from $\bar{K}$ or $\bar{K}'$ [see Figs.~\ref{fig:fig10}(b), \ref{fig:fig10}(d), \ref{fig:fig11}(b), and \ref{fig:fig11}(d)],
and additional peaks occur due to interband transitions from or to the ring of the crossed Dirac cones [marked as $K_{1}'$ in Fig.~\ref{fig:fig11}(d)] and due to interband transitions between other Dirac cones [marked as $K_{2}$ in Fig.~\ref{fig:fig10}(b) or $K_{3}$ and $K_{4}$ in Fig.~\ref{fig:fig11}(b)], which become significant for smaller $\theta$ or larger $U$.

\end{document}